\documentclass[aps,pra,showpacs,superscriptaddress,final,twocolumn]{revtex4}
\usepackage{graphicx}
\usepackage{times}
\usepackage{bm}
\usepackage{amstext}
\usepackage{amsfonts}
\usepackage{rotating}

\usepackage{psfrag}
\usepackage{color}

\definecolor{mygreen}{rgb}{0, 0.5, 0}

\begin{document}

\title{Quantum Scattering of Distinguishable Bosons using an Ultracold Atom Collider}
\author{Angela S. Mellish}\affiliation{Department of Physics, University of Otago,
Dunedin, New Zealand} \author{Niels Kj\ae
rgaard}\affiliation{Department of Physics, University of Otago,
Dunedin, New Zealand}\affiliation{Niels Bohr Institute, University
of Copenhagen, Denmark}\affiliation{QUANTOP---Danish National
Research Foundation Center for Quantum Optics}
\author{Paul S. Julienne}\affiliation{
National Institute of Standards and Technology, 100 Bureau Drive,
Stop 8423, Gaithersburg, Maryland, 20899-8423 USA}
\author{Andrew C.
Wilson}\affiliation{Department of Physics, University of Otago,
Dunedin, New Zealand}
\date{\today}
\begin{abstract}
We describe a new implementation of magnetic collider for
investigating cold collisions between ultracold atomic clouds in
different spin states, and we use this to investigate scattering
involving both even and odd order partial waves.  Our method relies
on the axial assymetry of a double-well magnetic trap to selectively
prepare the spin state in each cloud. We measure the energy
dependence of $s$, $p$ and $d$ partial wave phase shifts in
collisions up to 300~$\rm \mu$K between $\rm ^{87}Rb$ atoms in the
 $5S_{1/2}, F=1, m_F=-1$ and $5S_{1/2}, F=2, m_F=1$ states.
\end{abstract}


\maketitle Collisions in ultracold and degenerate quantum gases play
a key role in many of their interesting properties
\cite{Burnett2002}. So far, investigations with ultracold atoms have
been mostly concerned with {\textit s}-wave scattering processes,
but now nonzero partial waves play a critical role in many
investigations, (see, e.g., \cite{Regal2003}). A magnetic collider
scheme for determining the contribution made by higher-order partial
waves was recently implemented \cite{Thomas2004,Buggle2004}. In
these experiments the atoms were in the same spin state, limiting
the collisions to those involving only even-order partial waves ---
a consequence of the particles being indistinguishable bosonic
particles.

In the present work, we extend our collider method to
\textit{distinguishable} bosons for which the scattering is
fundamentally different since both odd and even angular momentum
components are allowed. As in our original work \cite{Thomas2004},
spin-polarized $^{87}$Rb atoms are loaded into a magnetic
double-well potential which is then transformed to a single well to
initiate a collision. Here, however, one of the clouds is converted
to a different spin state prior to collision making the scattering
patterns crucially different. We observe the interference of
{\textit s}, {\textit p} and {\textit d} partial waves for
collisions between atoms in the $F=1, m_F=-1$ and $F=2, m_F=1$
hyperfine ground states. Despite the complexity of the three-wave
interference, we successfully determine the three partial wave phase
shifts for energies up to 300~${\rm \mu}K$ as measured in units of
the Boltzmann constant $k_B$.

The angular dependence of the two-body scattering problem is
described by the complex scattering amplitude $f(\theta)$
\cite{Taylor1972}. Using the partial wave expansion, this is
expressed as
$f(\theta)=\frac{1}{2ik}\sum_{\ell=0}^{\infty}(2\ell+1)(e^{2i\eta_\ell}-1)P_\ell(\cos\theta)$,
where $P_\ell$ is the $\ell^{\rm th}$ order Legendre polynomial and
$\eta_\ell$ are the partial wave phase shifts which depend on the
scattering potential and relative wave vector $k$ of the colliding
atom pair. For the range of energies we focus on here, only the
first three partial waves $\ell=0,1,2$ contribute \cite{note1}. In
this case the differential cross-section $d\sigma/d\Omega =
|f(\theta)|^2$ is given by
\begin{widetext}
\begin{eqnarray}\label{diffcross}
\frac{{\rm d }\sigma}{{\rm d}\Omega}&=&\frac{1}{k^2}\{\sin^2\eta_0 +
9\sin^2\eta_1\cos^2\theta +
\frac{25}{4}\sin^2\eta_2(3\cos^2\theta-1)^2  +
6\sin\eta_0\sin\eta_1\cos(\eta_0-\eta_1)\cos\theta \nonumber \\  &+&
5\sin\eta_0\sin\eta_2\cos(\eta_0-\eta_2)(3\cos^2\theta-1) +
15\sin\eta_1\sin\eta_2\cos(\eta_1-\eta_2)(3\cos^2\theta-1)\cos\theta\}.
\end{eqnarray}
\end{widetext}
Because of the orthogonality and completeness of the Legendre
polynomials, a fit of an interference expression in the form
Eq.~(\ref{diffcross}) to a measured angular distribution directly
gives the partial wave phase shifts $\eta_0$, $\eta_1$ and $\eta_2$
irrespective of knowledge about absolute quantities such as particle
flux \cite{Buggle2004}.

Our experimental procedure is as follows. $^{87}$Rb atoms in the
$5S_{1/2} F=1, m_F=-1$ $(\equiv|1\rangle)$ state are loaded into a
magnetic quadrupole-Ioffe-configuration (QUIC) trap
\cite{Esslinger1998} with trap frequencies $\omega_z/2\pi= 11$~Hz
axially and $\omega_\rho/2\pi= 90$~Hz radially. The details of
loading the double-well trap and initiating a collision are much the
same as described in~\cite{Thomas2004}. In summary, after rf-induced
evaporation of the atoms to a temperature of approximately $2~\rm
\mu$K we adiabatically transform the potential to a double well by
raising a potential barrier along the axial dimension of the trap to
split the cloud in half \cite{Thomas2002}. The clouds are then
further evaporatively cooled to a temperature of typically a few
hundred nano-Kelvin, just above the Bose-Einstein condensation
transition temperature. A collision between the clouds is initiated
by rapidly transforming the potential back to a single well. The
collision energy is selected by adjusting the well spacing in the
double-well trap.

To enable a collision between atoms in different spin states, we
apply a two-photon pulse consisting of a microwave ($\sim6.8$~GHz)
and an rf ($\sim2$~MHz) photon (depending on the Zeeman splitting)
to transfer $|1\rangle$ state atoms to the $5S_{1/2}, F=2, m_F=1$
$(\equiv|2\rangle)$ state \cite{Matthews1998}. Due to the intrinsic
axial asymmetry of the QUIC trap the clouds are situated at slightly
different magnetic field values immediately after the double- to
single-well trap transformation. This enables us to selectively
address and convert up to 90~$\%$ of the atoms in one of the clouds,
while only 10~$\%$ of the atoms in the other cloud are converted to
the $|2\rangle$ state. To first order, the $|1\rangle$ and
$|2\rangle$ states have the same magnetic moment and experience the
same confinement potential.

\begin{figure}
\includegraphics[width=0.8\columnwidth]{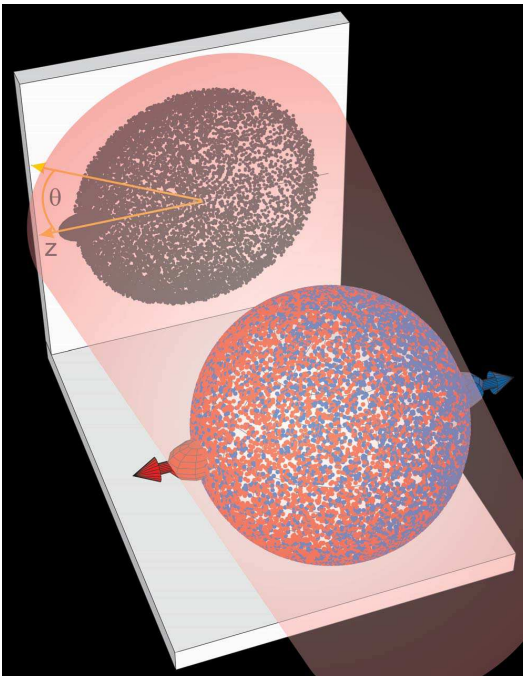}
\caption{(color online) After a collision between two atomic clouds
in different spin states, pairs of diametrically-opposite scattered
particles will be distributed over the expanding Newton sphere
according to the differential cross-section. Using a light beam
resonant with only one of the states (depicted as red), an
absorption image of the contribution of this particular state to the
scattering halo is obtained.} \label{imaging}
\end{figure}
To selectively probe the scattered $|2\rangle$ state atoms we apply
a $\rm 20~\mu s$ pulse of resonant light on the $5S_{1/2}, F=2
\rightarrow 5P_{3/2}, F'=3$ transition along a radial direction
shortly after the end of the collision, and acquire an absorption
image. This leaves the $|1\rangle$ state atoms undetected. An
illustration of this is shown in Fig.~\ref{imaging}. Alternatively,
we can simultaneously probe both the $|1\rangle$ and $|2\rangle$
state atoms by applying some $5S_{1/2}, F=1 \rightarrow 5P_{3/2},
F'=2$ light to pump all of the atoms to the $F=2$ level shortly
before the probing pulse.

\begin{figure}
\includegraphics[width=0.7\linewidth]{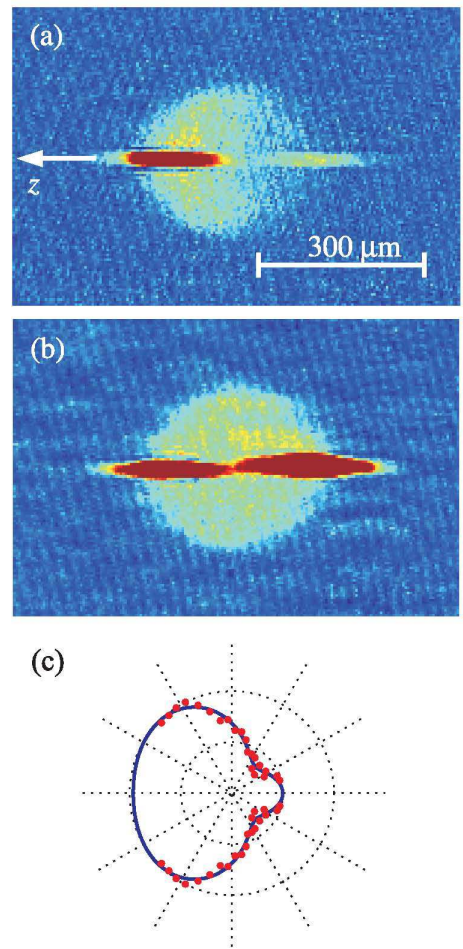}\\
\caption{(color online) Scattering images for a collision at
$E/k_B=135~{\rm \mu}$K, (a) probing only atoms in the $F=2$ state
and (b) probing both the $F=1$ and $F=2$ states. The asymmetry in
the scattering pattern of (a) is due to {\textit p}-wave scattering.
The corresponding angular scattering probability is shown in (c)
with a fit to Eq.~(\ref{diffcross}) (solid line).} \label{pwave}
\end{figure}

\begin{figure}
\includegraphics[width=0.6\linewidth]{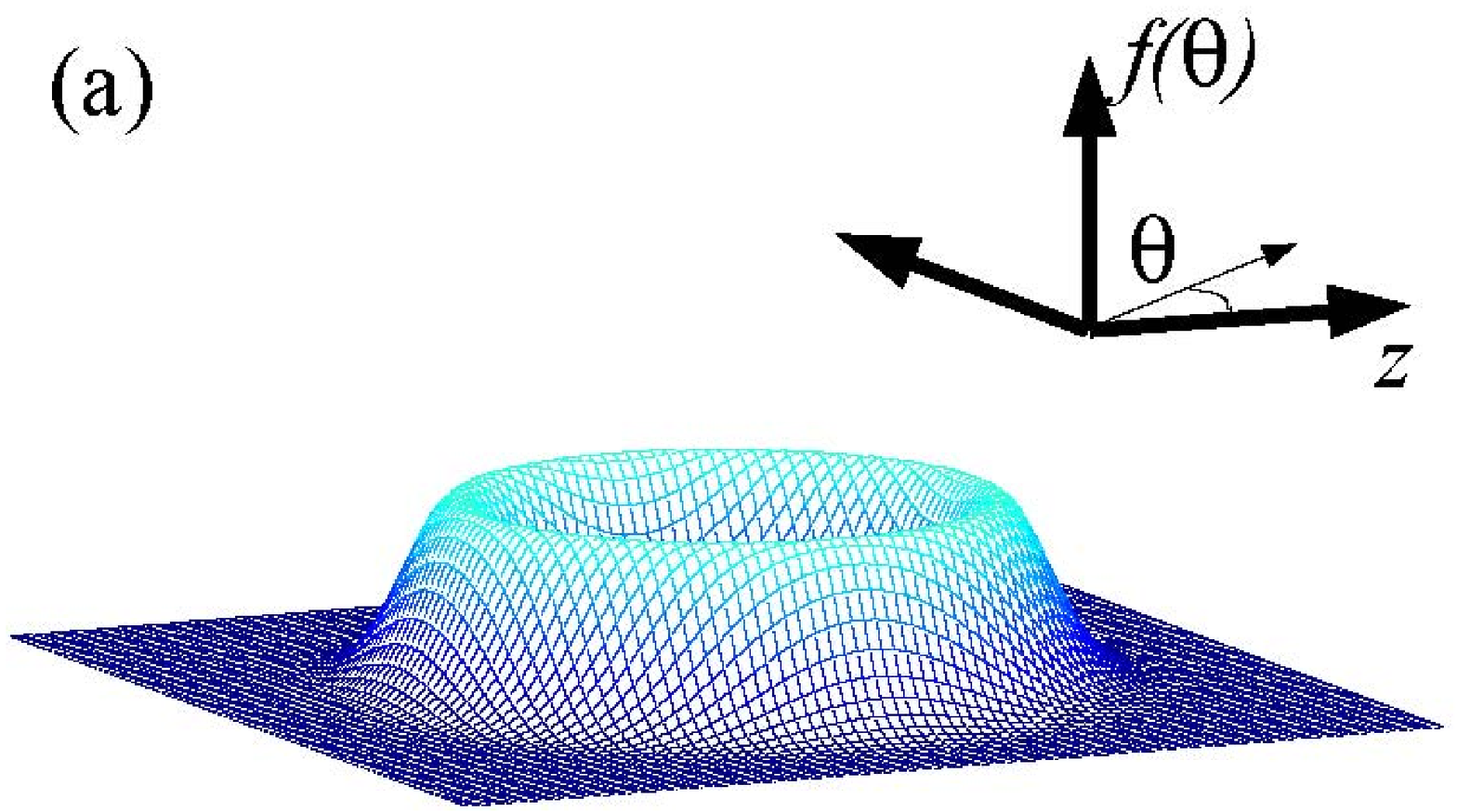}\\
\vspace{0.1cm}
\includegraphics[width=0.6\linewidth]{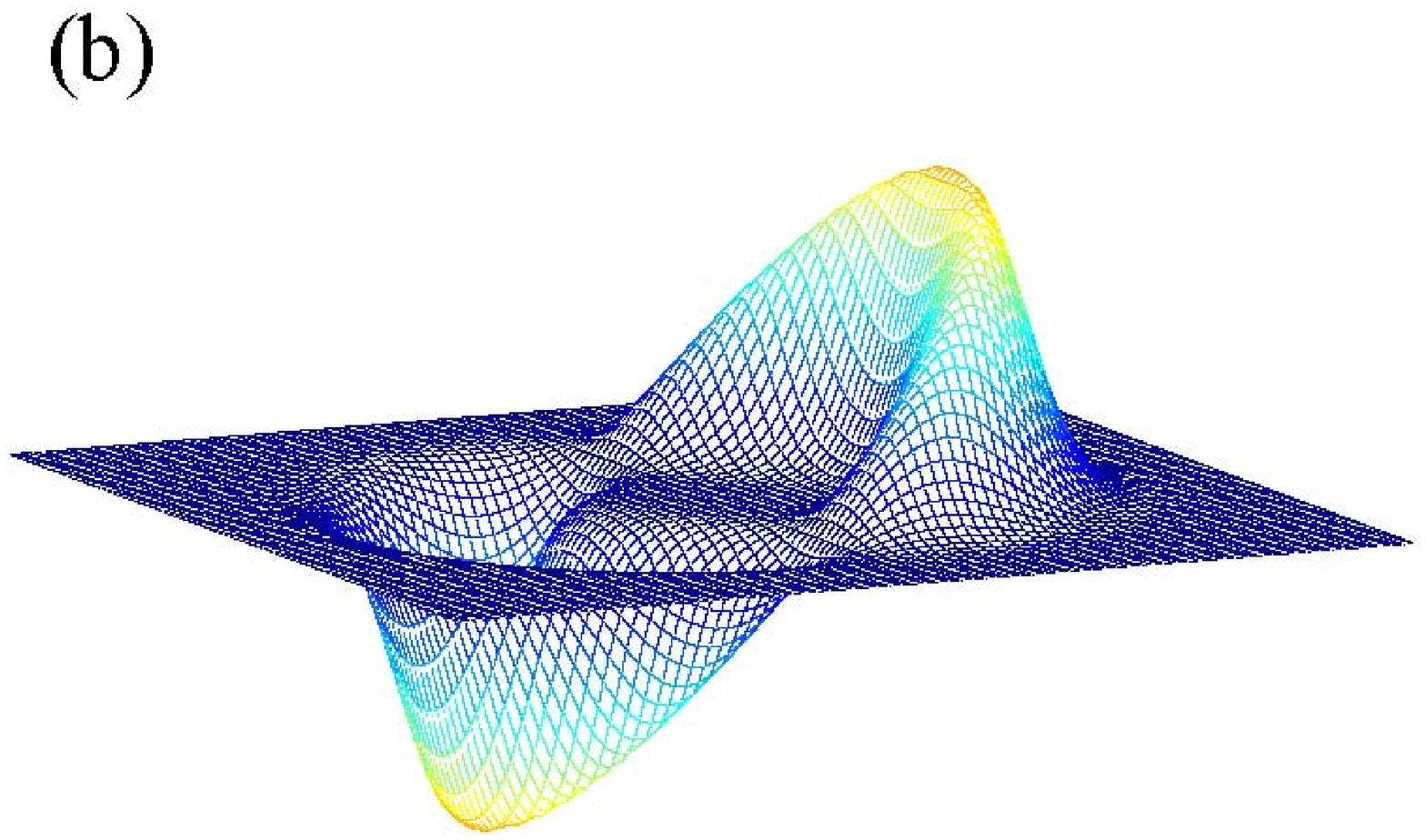}\\
\vspace{0.1cm}
\includegraphics[width=0.6\linewidth]{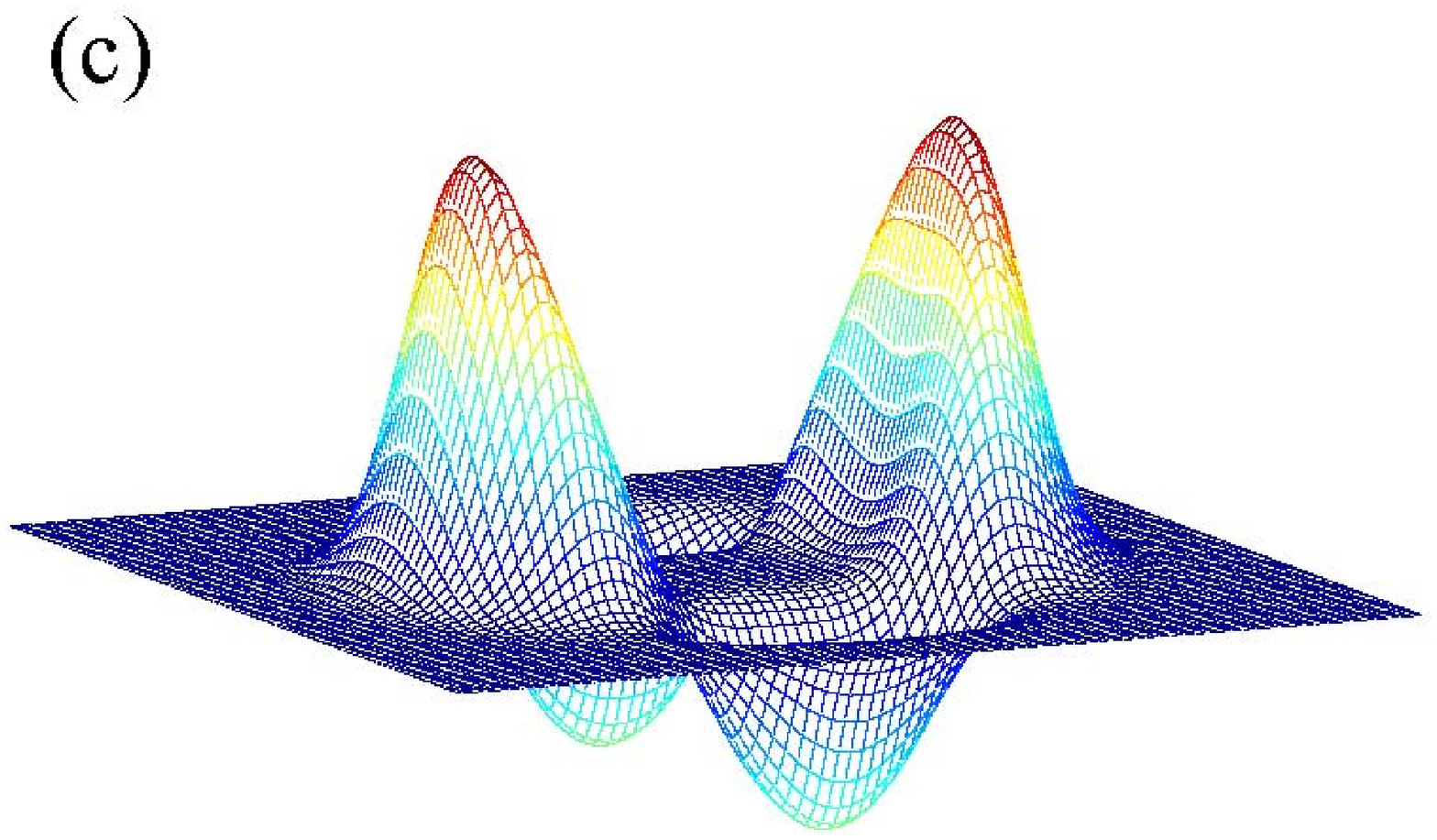}
\caption{(color online) A graphical representation of the
contributions to the scattering amplitude for the first three
partial waves; (a) {\textit s}-wave, (b) {\textit p}-wave and (c)
{\textit d}-wave. The sign and magnitude of each $\ell$ term in
$f(\theta)$ is plotted vertically for a spherical scattering shell
with a Gaussian profile in the radial direction. The relative scale
of each is determined by $\eta_\ell$. In contrast to {\textit s}-
and {\textit d}-wave, the {\textit p}-wave contribution to
$f(\theta)$ is antisymmetric in $\theta$.} \label{partialwaves}
\end{figure}

Figure~\ref{pwave} shows absorption images after a collision
at $E/k_B=135~{\rm \mu}$K between atomic clouds in the $|1\rangle$
and $|2\rangle$ states.
In Fig.~\ref{pwave}(a) only atoms in the $|2\rangle$ state have been
probed, whereas in (b) atoms in both the $|1\rangle$ and $|2\rangle$
states are imaged. The distinct left-right asymmetry of the
scattered atoms in (a) is the result of partial-wave interference
between the odd ($\ell=1$) {\textit p}-wave and even {\textit s}-
and {\textit d}-waves. The scattering amplitude of the {\textit
p}-wave component changes sign at $\theta=\pm\pi/2$ as can be seen
in Fig.~\ref{partialwaves}. For the collision energy in this
example, where the {\textit d}-wave contribution is relatively
small, the {\textit p}-wave interferes constructively with {\textit
s}-wave for angles $|\theta|<\pi/2$ and destructively for
$|\theta|>\pi/2$ where $\theta$ is defined with respect to the
collision axis in the initial direction of travel (\textit{i.e.,}
for the $|2\rangle$ state shown in Fig.~\ref{pwave},
$|\theta|<\pi/2$ is to the left of the image). Since $\theta$ is
defined with the opposite sense for the $|1\rangle$ and $|2\rangle$
states, $f(\theta)$ for the $|1\rangle$ state is complementary to
that of $|2\rangle$ and imaging both states together results in a
symmetric scattering pattern [Fig.~\ref{pwave}(b)].

\begin{figure}
\centering
\includegraphics[width=\linewidth]{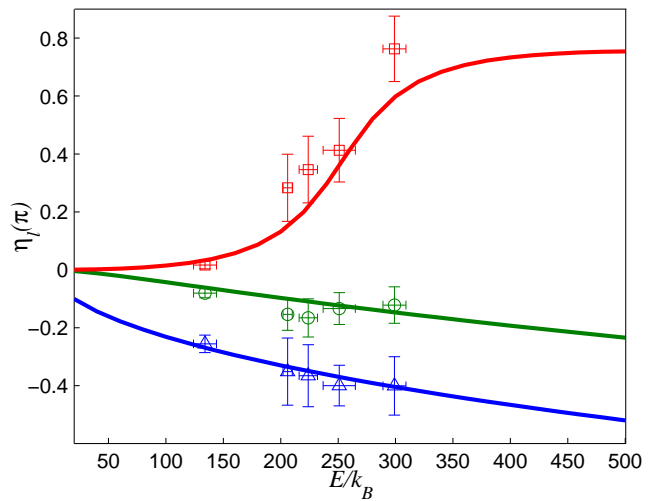}
\caption{(color online) The partial wave phase shifts for collisions
between the $|1\rangle$ and $|2\rangle$ states. The symbols
represent the {\it s} (\textcolor{blue}{$\triangle$}), {\it p}
(\textcolor{mygreen}{$\bigcirc$}) and {\it d}
(\textcolor{red}{$\Box$}) phase shifts extracted from the data and
the solid lines are a theoretical calculation from a
coupled-channels model.} \label{dataplot}
\end{figure}

We analyze the absorption images of the scattering patterns using
the method described in~\cite{Kjaergaard2004}. Briefly, we
reconstruct the 3D distribution of the scattered atoms using the
inverse Abel transformation \cite{Dribinski}. The Abel-inverted
image is divided into 30 angular bins which reflect the trajectories
of scattered atoms in the harmonic potential. The number of
scattered particles in each of the bins yields a measure of the
angular scattering probability, which is proportional to the
differential cross-section in Eq.~(\ref{diffcross}). We fit
Eq.~(\ref{diffcross}) to this data to obtain the partial wave phase
shifts $\eta_0$, $\eta_1$ and $\eta_2$ for the {\textit s}, {\textit
p}, and {\textit d} partial waves respectively. As emphasized by
Buggle \textit{et al.\ }\cite{Buggle2004}, this is an
interferometric method which does not rely on absolute particle
numbers and identifies only the amplitudes and relative signs of the
phase shifts. The {\textit s}-wave scattering length is known to be
positive (repulsive interaction) for the states considered here so
we choose the corresponding solution where $\eta_0<0$ for our energy
range. The collision energy $E=mv_{\rm rel}^2/4=\hbar k^2/m$ is
measured within a typical uncertainty of 5~$\mu K$ by determining
the relative velocity $v_{\rm rel}$ from a linear fit to the
position of the clouds over approximately 2~ms either side of
collision. In Fig.~\ref{dataplot} each phase shift value is the
average of up to 10 measurements at the particular collision energy.
The error bars on the data combine statistical uncertainty and
errors associated with the fit to Eq.~(\ref{diffcross}).

A comparison of the measurements to theoretical predictions is shown
in Fig.~\ref{dataplot}. These are standard coupled-channels
numerical calculations \cite{Stoof1988,Mies1996,Julienne2002} for
the collision of two atoms in hyperfine states $F,M$ and $F',M'$ in
a low magnetic field $B$ with relative (partial wave) angular
momentum $\ell$ and projection $m$. All channels $\{FM,F'M',\ell
m\}$ coupled by terms in the molecular Hamiltonian are included.
Only channels with $M_\mathrm{tot}=M+M'+m$ can couple to one another
and because the collisions are from a single direction (defined by
the vector connecting the two initial separated atomic clouds) we
need only include the $M_\mathrm{tot}=0$ channels. The Hamiltonian
contains the radial $T_R$ and rotational $T_\mathrm{rot}$ kinetic
energy terms, the electron-electron spin-spin interaction $\alpha^2
H_{\rm ss}$ (where $\alpha$ is the fine structure constant), the
electron-nuclear spin interaction terms $\alpha^2 H_{\rm hf}$ that
gives the atomic hyperfine energies, and the strong chemical
interactions described by the two adiabatic Born-Oppenheimer
potential curves that correlate with two separated $^2$S atoms.
These potential curves correspond to the electronic states of
$^1\Sigma_g^+$ and $^3\Sigma_u^+$ symmetry. There are 8 $s$-wave
channels needed to describe $M_\mathrm{tot}=0$ $s$-wave collisions
of $|2\rangle$ and $|1\rangle$ atoms (5 open and 3 closed). There
are also 18 $M_\mathrm{tot}=0$ $p$-wave collision channels (11 open
and 7 closed), and 30 $M_\mathrm{tot}=0$ $d$-wave channels (18 open
and 12 closed). All of these channels are included in the basis set
for each partial wave. If the channels are designated by the index
$j$, so that the wavefunction for atoms in the entrance channel $i$
is $\Psi_i=\sum_j |j\rangle f_{ji}(R)/R$, the coupled Schr\"{o}dinger
equations, in a basis set defined by the separated atom quantum
numbers, takes on the form
\parbox{\linewidth}{
\begin{eqnarray}
 \frac{\hbar^2}{2\mu} \frac{d^2f_{ki}}{dR^2} + \left (E-E_k  - \frac{\hbar^2 \ell_k (\ell_k+1)}{2 \mu R^2} \right ) f_{ki}(R)  \nonumber \\
  - \sum_j V_{kj}(R)f_{ji}(R) = 0.
\end{eqnarray}
} Here $E_k$ and $\ell_k$ are the respective Zeeman energy and
relative angular momentum quantum number of the two colliding
separated atoms for the magnetic field $B$, and the potential matrix
elements $V_{kj}$ define the interchannel coupling. These equations
are solved numerically using standard algorithms \cite{gordon1969}.
For comparison with the data, the calculation uses a magnetic field
of 0.23 mT, and the scattering potentials are characterized by a
dispersion coefficient $C_6 = 4703$~au and triplet $a_t =
+98.96$~a$_0$ and singlet $a_s = +90.1$~$a_0$ scattering lengths
consistent with \cite{vanKempen2002} (1 au $=E_h a_0^6$, where $E_h
= 4.36 \times 10^{-18}$ J and a$_0=$ 0.0529 nm).

As can be seen in Fig.~\ref{dataplot}, our experimental observations
are described well by the theoretical model. The dramatic change of
the {\textit d}-wave phase shift is a signature for the {\textit
d}-wave shape resonance known to occur for collisions between two
$^{87}$Rb atoms \cite{Thomas2004,Buggle2004,Boesten97}. We estimate
the position of the resonance to be $(235\pm50)~\mu$K with a width
of approximately $120~\mu$K (FWHM) from a Lorentzian fit to the data
around the resonance. Calculated inelastic collision rate constants
remain below $10^{-13}$ cm$^3/$s over the collision range of
interest (compared to a maximum total elastic scattering
cross-section of $\sim 1.6\times 10^{-11}$~cm$^2$), even when
enhanced by the $d$-wave shape resonance. This is due to the
exceptional case that both potentials have similar scattering phase
shifts at low collision energies for threshold $^{87}$Rb
spin-exchange
relaxation~\cite{Burke1997,Julienne1997,Kokkelmans1997}.
Correspondingly, we do not observe any atom loss from the trap
resulting from the collision.

Two effects are not included in our analysis: state impurities in
the clouds and the possibility of multiple scattering. The first of
these is a difficult technical issue relating to our set-up and the
second is of a more fundamental nature. With state impurities in
both clouds, the collision processes which can occur are
$|1\rangle+|2\rangle$, $|2\rangle+|2\rangle$, $|1\rangle+|1\rangle$,
and $|2\rangle+|1\rangle$, with relative amounts depending on the
density of impurities. If these effects were significant one would
expect the presence of collisions between the $|1\rangle$ and
$|2\rangle$ states in the ``wrong'' direction to diminish the
measured {\textit p}-wave contribution, whereas scattering due to
the $|2\rangle+|2\rangle$ and $|1\rangle+|1\rangle$ collision
processes would increase the perceived {\textit s}- and {\textit
d}-wave phase shifts measured which is clearly not the case in
Fig.~\ref{dataplot}. As for the second issue, we observe only
approximately one-third of the total number of atoms scattered after
a collision near the resonance, indicating that the probability of a
secondary collision is relatively small. A detailed theoretical
analysis of multiple scattering is difficult outside the {\textit
s}-wave regime, and particularly near a {\textit d}-wave shape
resonance, since the energy and centre-of-mass of a subsequent
collision depend crucially on the scattering angle after the first
collision.

In conclusion, we have investigated the energy dependence of
collisions between two $^{87}$Rb clouds in different spin states.
Our experimental observations agree well with predictions from a
theoretical coupled-channels model. We note that the collision
between two such particles of different spins provides a mechanism
for producing spin entanglement. The resulting pair correlation
could potentially be observed as in recent experiments on
dissociating molecules \cite{Greiner2005} and colliding
Bose-Einstein condensates \cite{westbrook2006}. Furthermore, the
occurrence of a {\textit d}-wave resonance and the resulting
directionality of scattered particles may serve as a vehicle for the
production of pair correlated beams.

This work has been partially supported by the Marsden Fund of the
Royal Society of New Zealand (grant 02UOO080) and the U.\ S.\ Office
of Naval Research. N.\ K.\ acknowledges the support of the Danish
Natural Science Research Council.

\newcommand{\Ch}{Ch}


\begin{thebibliography}{30}
\bibitem{note1}Calculations show that contributions from higher
partial wave are negligible, e.g., $\eta_4<0.02\times\pi$ for
collision energies up to 500~$\rm \mu K$.
\end{thebibliography}
\end{document}